\documentclass[acmsmall]{acmart}

\setcopyright{acmcopyright}
\acmJournal{PACMHCI}
\acmYear{2022} \acmVolume{6} \acmNumber{CSCW2} \acmArticle{266} \acmMonth{11} \acmPrice{15.00}\acmDOI{10.1145/3555156}
\usepackage{caption}
\usepackage{booktabs}
\usepackage{subcaption}
\usepackage{graphicx}
\usepackage[export]{adjustbox}
\AtBeginDocument{%
  \providecommand\BibTeX{{%
    \normalfont B\kern-0.5em{\scshape i\kern-0.25em b}\kern-0.8em\TeX}}}

\newcommand{\ndcg}{
\begin{table}[]
\centering
\small
\begin{tabular}{@{\hspace{0cm}}lllll@{}}
\toprule
\textbf{Rank}  & \textbf{PageRank} & \textbf{Pageviews} & \textbf{Edit Count} & \textbf{Vital Articles} \\ \midrule
1 & Countries & Women & Countries & Anthropology \\
2 & Cities & Film & Rock music & Philosophy \\
3 & Former Countries & Countries & Animation & History of Science \\
4 & Higher education & United States & Pop music & Science \\
5 & UK geography & Comedy & Disney & Sociology \\
6 & Politics & Television & Professional wrestling & History \\
7 & Geography & California & Skepticism & Religion\\
8 & Ethnic groups & Pop music & Football & Environment \\
9 & Politics of the United Kingdom & Biography & Metal & Tree of Life \\
10 & Companies & NBA & Conservatism & Animals \\ \bottomrule
\end{tabular}
\caption{Top 10 WikiProjects by normalized discounted cumulative gain (nDCG) for each prioritization method.}
\label{tab:ndcg}
\end{table}
}

\newcommand{\criteria}{
\begin{table}[]
\centering
\small
\resizebox{\textwidth}{!}{
\begin{tabular}{@{\hspace{0cm}}ll@{}}
\toprule
\textbf{Importance Criterion} & \textbf{Example Quote} \\ \midrule
Everyday Significance & \textit{``An activity {[}sleep{]} that takes up 1/3 of your lifetime seems to be pretty vital to me.''} \\
Cultural Significance & \textit{``Sports have in some form been a part of the vast majority of cultures for much of there history.''} \\
Historical Significance & \textit{``The concept {[}bourgeoisie{]} has had a massive role in human history.''} \\
Enduring Significance & \textit{``The repercussions {[}of the 2019-20 coronavirus pandemic{]} will be felt for many decades, at the very least.''} \\
Breadth & \textit{``Folklore is the broader and more fundamental article {[}compared to Myth{]}.''} \\ \midrule
\textbf{Global Criterion} & \textbf{Example Quote} \\ \midrule
Balance & \textit{``If sport receives enough support then I think we should add an almost equivalent female dominated activity to balance things out (maybe dance).''} \\
Non-redundancy & \textit{``Everything on Earth is covered by Earth, and everything beyond Earth is of interest pretty much only for astronomy, which is covered by Science.''} \\
Completeness & \textit{``The only type of activism we lack is women’s rights - of which i would support Emmeline Pankhurst.''} \\ \bottomrule
\end{tabular}
}
\caption{Prioritization criteria derived from qualitative analysis for RQ1 with example quotes from VA discussions.}
\label{tab:criteria}
\end{table}
}

\newcommand{\vadiscussion}{
\begin{figure}
    \centering
    \includegraphics[width=\textwidth, frame]{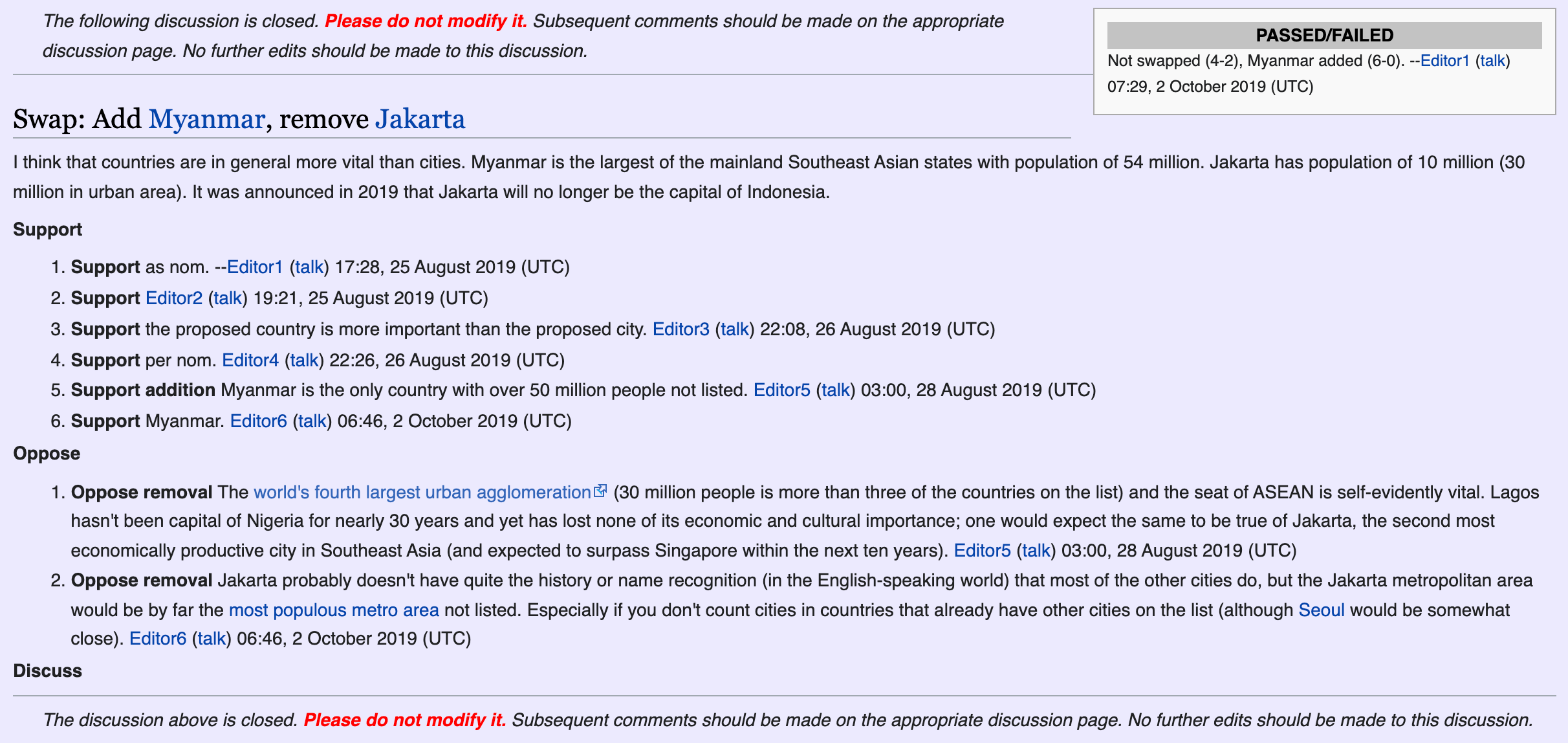}
    \caption{Example of a typical discussion on the VA talk pages. Editors' usernames have been replaced with numbered placeholders.}
    \label{fig:vadiscussion}
\end{figure}
}
\newcommand{\rqthreegender}{
\begin{figure}
    \centering
    \includegraphics[width=.9\textwidth, frame]{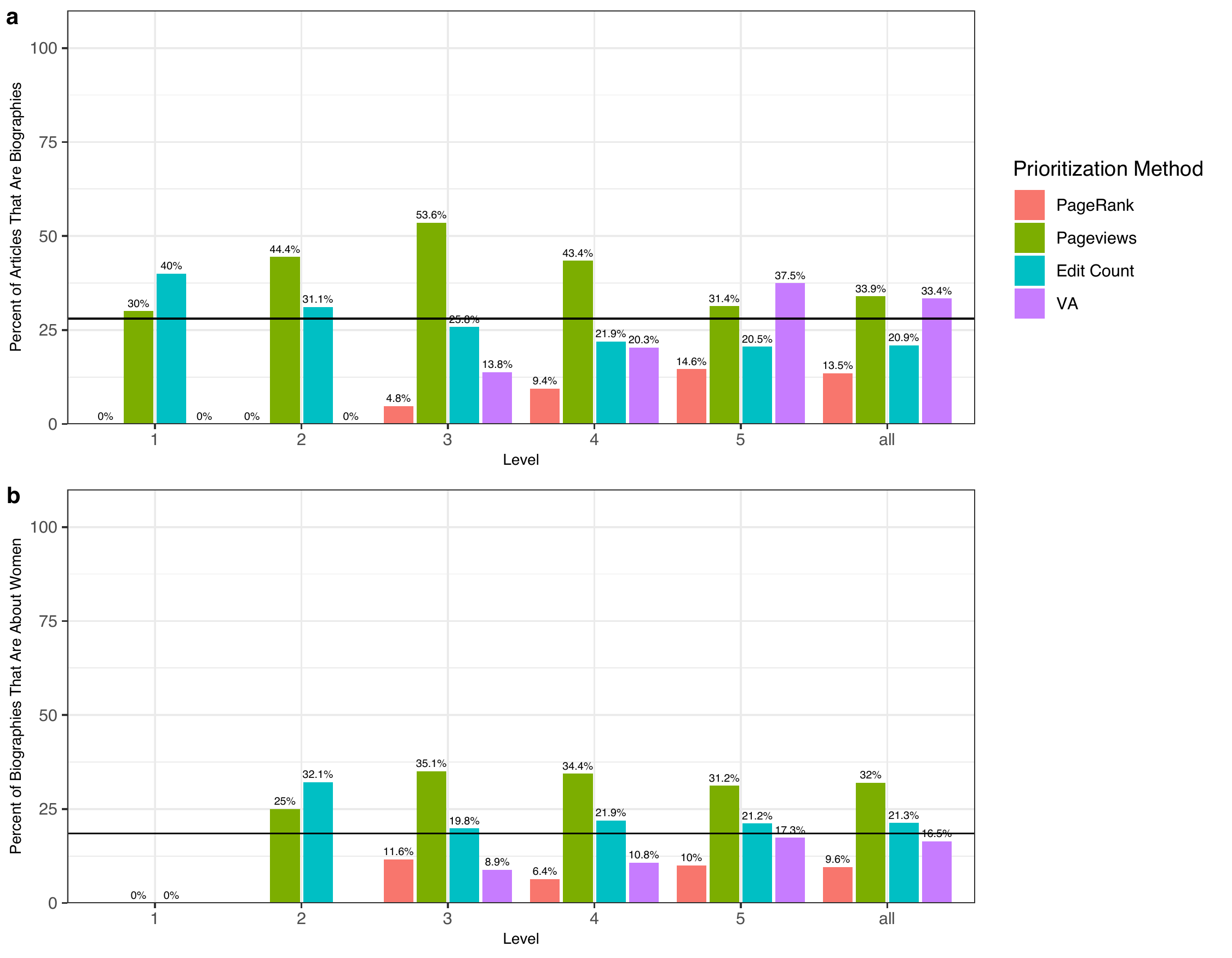}
    \caption{Gender breakdown of each prioritization method by level. The horizontal line shows the corresponding percentage on English Wikipedia as a whole.}
    \label{fig:rqthreegender}
\end{figure}
}

\newcommand{\rqthreegeo}{
\begin{figure}
    \centering
    \includegraphics[width=.9\textwidth, frame]{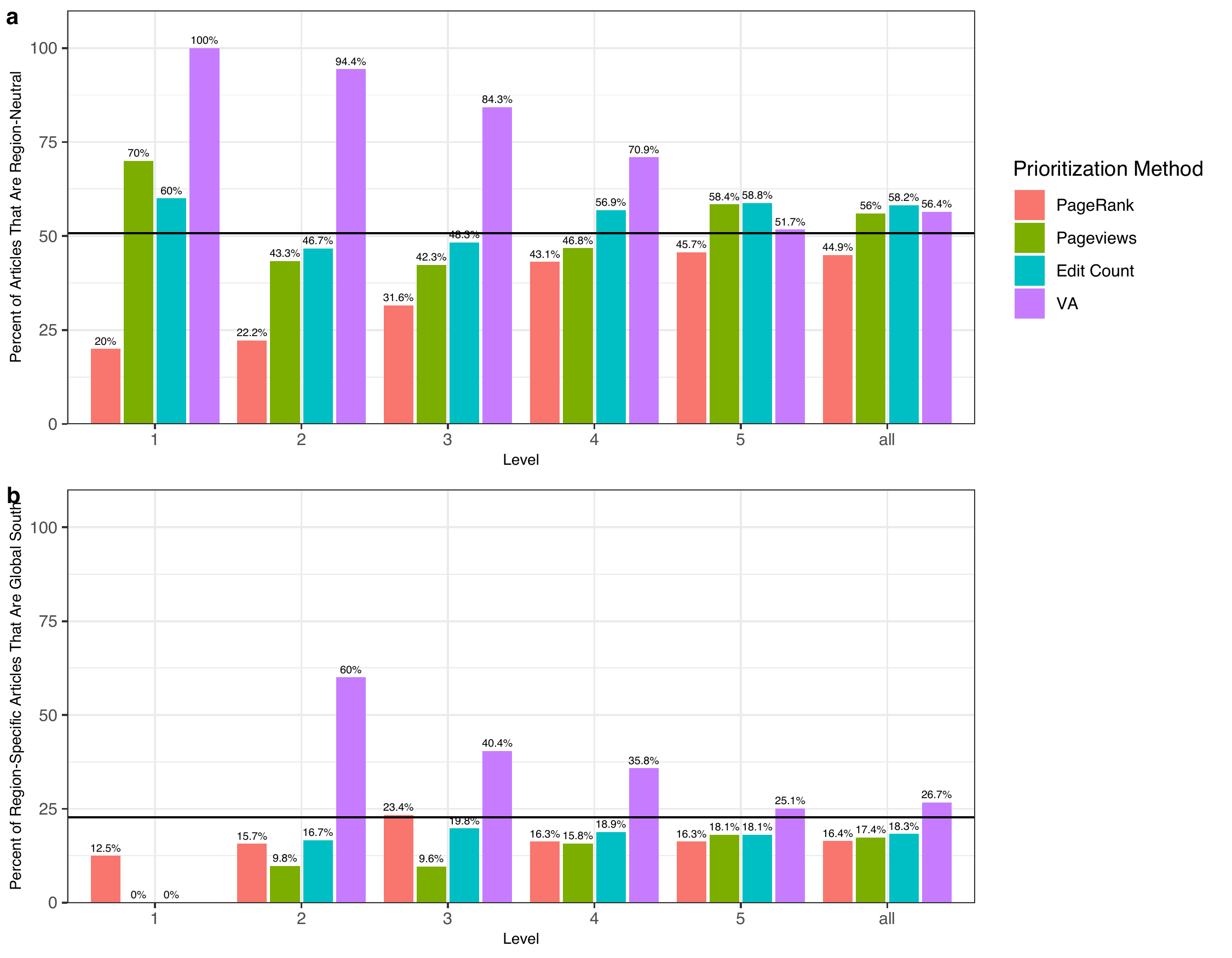}
    \caption{Geographical breakdown of each prioritization method by level. The horizontal line shows the corresponding percentage on English Wikipedia as a whole.}
    \label{fig:rqthreegeo}
\end{figure}
}

\newcommand{\rqfourgender}{
\begin{figure}
\centering
\begin{subfigure}{\textwidth}
  \centering
  \includegraphics[width=.9\linewidth]{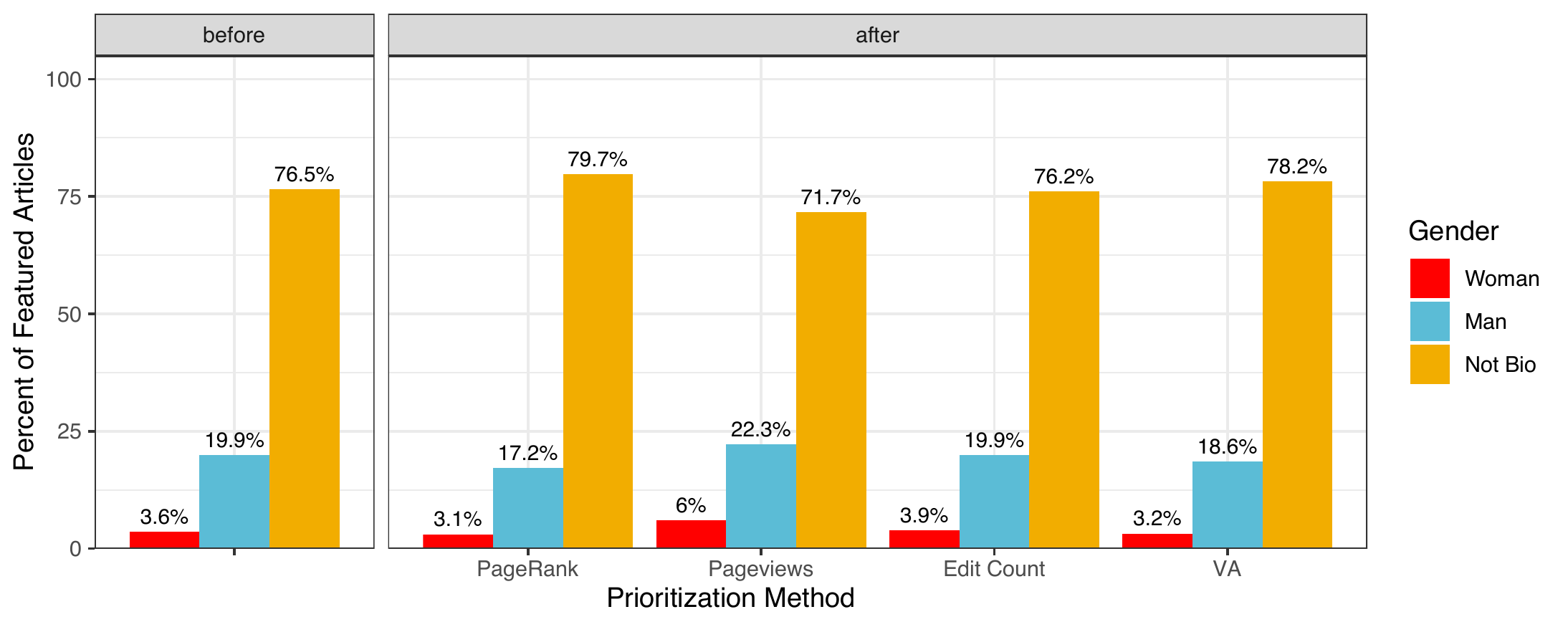}
  \label{fig:rqfourgender1}
\end{subfigure}
\begin{subfigure}{\textwidth}
  \centering
  \includegraphics[width=.9\linewidth]{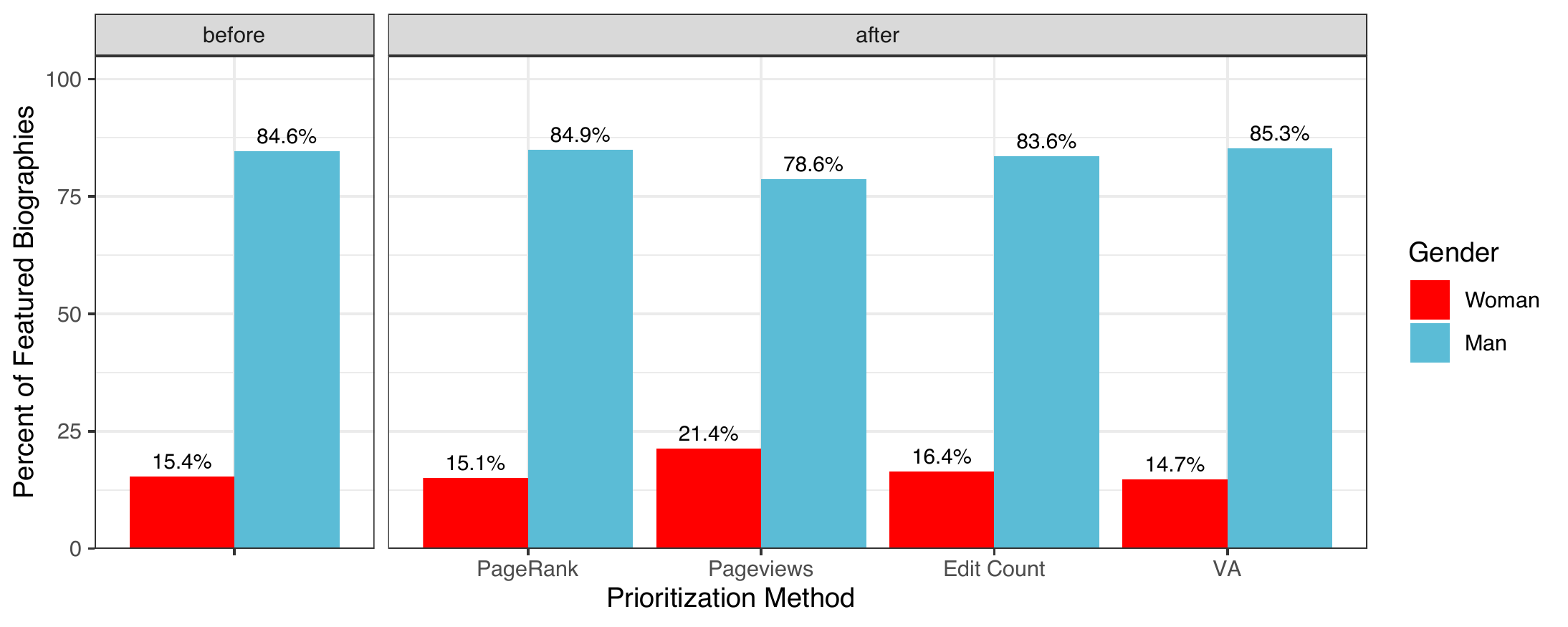}
  \label{fig:rqfourgender2}
\end{subfigure}
\caption{Gender composition of Featured articles before and after perfect prioritization over 5 years using each method.}
\label{fig:rqfourgender}
\end{figure}
}

\newcommand{\rqfourgeo}{
\begin{figure}
\centering
\begin{subfigure}{\textwidth}
  \centering
  \includegraphics[width=.9\linewidth]{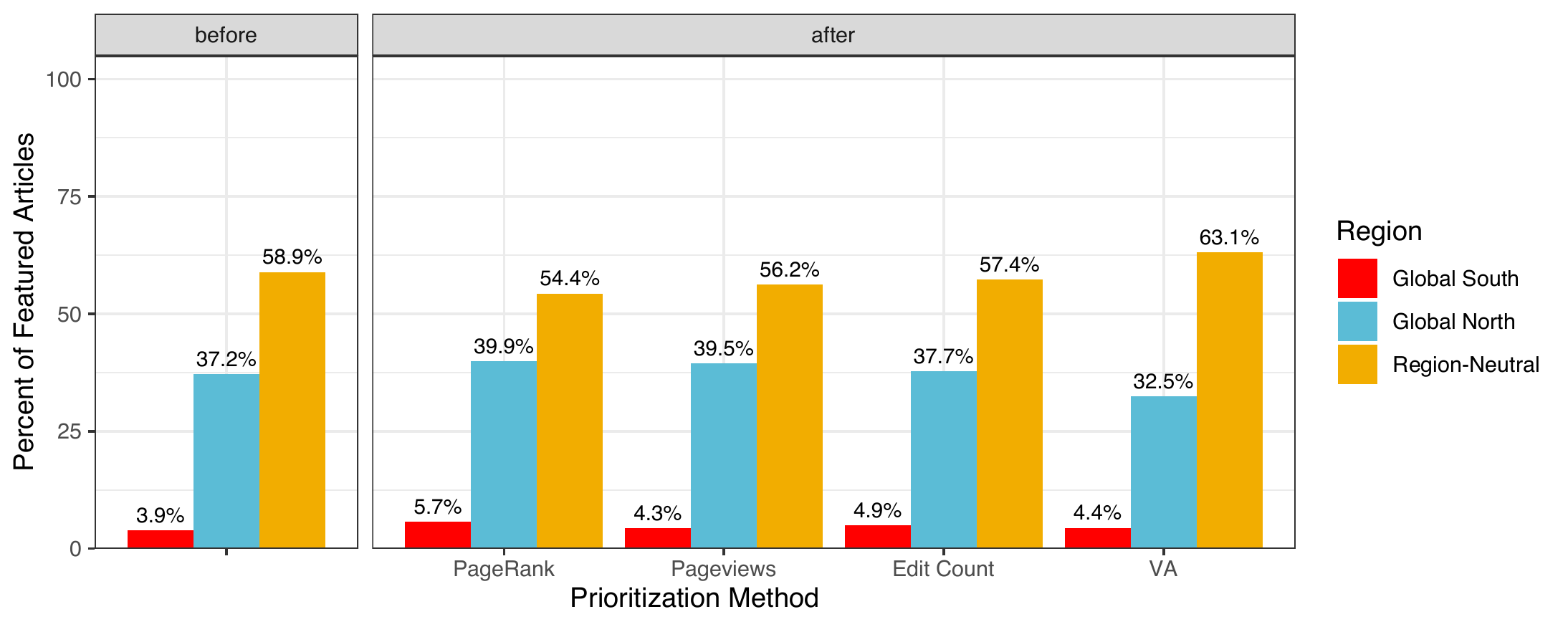}
  \label{fig:rqfourgeo1}
\end{subfigure}
\begin{subfigure}{\textwidth}
  \centering
  \includegraphics[width=.9\linewidth]{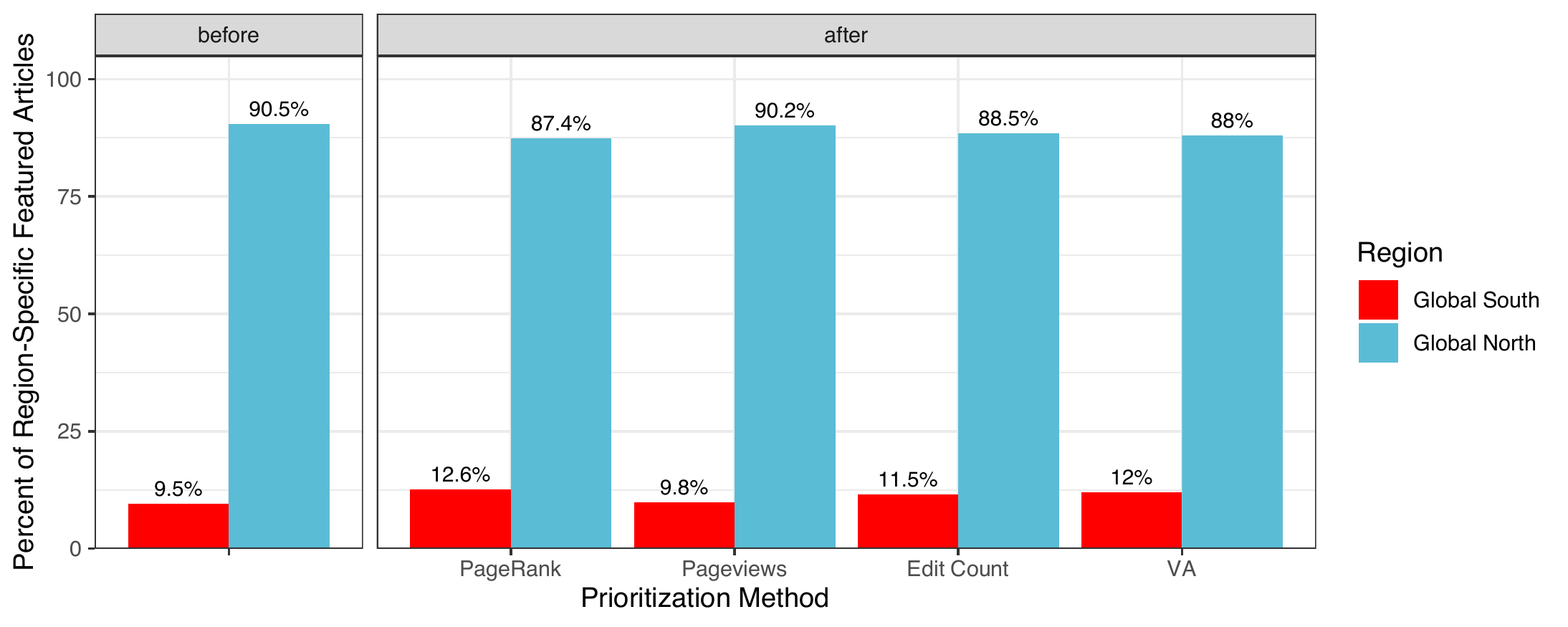}
  \label{fig:rqfourgeo2}
\end{subfigure}
\caption{Geographical composition of Featured articles before and after perfect prioritization over 5 years using each method.}
\label{fig:rqfourgeo}
\end{figure}
}

\usepackage{xcolor}

\newcommand\hl[1]{%
  \bgroup
  \hskip0pt\color{black}%
  #1%
  \egroup
}

\begin{document}

\title{``We Need a Woman in Music'':\\Exploring Wikipedia’s Values on Article Priority}

\author{Mo Houtti}
\email{houtt001@umn.edu}
\affiliation{
  \institution{University of Minnesota}
  \city{Minneapolis}
  \state{Minnesota}
  \country{USA}
}
\author{Isaac Johnson}
\email{isaac@wikimedia.org}
\affiliation{
  \institution{Wikimedia Foundation}
  \country{USA}
}
\author{Joel Cepeda}
\email{joelacepeda@gmail.com}
\affiliation{
  \institution{Reading Plus}
  \country{USA}
}
\author{Soumya Khandelwal}
\email{khand080@umn.edu}
\affiliation{
  \institution{University of Minnesota}
  \country{USA}
}
\author{Aviral Bhatnagar}
\email{bhatn042@umn.edu}
\affiliation{
  \institution{University of Minnesota}
  \country{USA}
}
\author{Loren Terveen}
\email{terveen@umn.edu}
\affiliation{
  \institution{University of Minnesota}
  \country{USA}
}


\begin{abstract}
  Wikipedia---like most peer production communities---suffers from a basic problem: the amount of work that needs to be done (articles to be created and improved) exceeds the available resources (editor effort). Recommender systems have been deployed to address this problem, but they have tended to recommend work tasks that match individuals’ personal interests, ignoring more global community values. In English Wikipedia, discussion about \emph{Vital articles} constitutes a proxy for community values about the types of articles that are most important, and should therefore be prioritized for improvement. We first analyzed these discussions, finding that an article’s priority is considered a function of 1) its \emph{inherent importance} and 2) its effects on Wikipedia’s \emph{global composition}. One important example of the second consideration is balance, including along the dimensions of gender and geography. We then conducted a quantitative analysis evaluating how four different article prioritization methods\hl{---two from prior research---}would affect Wikipedia’s overall balance on these two dimensions; we found significant differences among the methods. We discuss the implications of our results, including particularly how they can guide the design of recommender systems that take into account community values, not just individuals' interests.
\end{abstract}

\ccsdesc[500]{Human-centered computing~Empirical studies in HCI}
\ccsdesc[500]{Human-centered computing~Wikis}

\keywords{Wikipedia; recommender systems; peer production; mixed methods}

\maketitle

\section{Introduction}

English Wikipedia’s About page describes the platform as \emph{``an online free-content encyclopedia project helping to create a world in which everyone can freely share in the sum of all knowledge''}~\cite{noauthor_wikipediaabout_2021}. Making the sum of all knowledge available is a massive undertaking; the English version of Wikipedia alone currently contains over 6 million articles. And despite having a large editor base that collectively makes approximately 5 million edits every month~\cite{noauthor_wikipediastatistics_2021}, much content on the platform is still of lower quality than would be desired; as of July 2020, only 5822 of English Wikipedia’s 6 million+ articles have been assigned the highest quality rating possible~\cite{noauthor_wikipediafeatured_2020}.

It is therefore useful to examine Wikipedia through the lens of the economizing problem~\cite{intriligator_economizing_2002}, whereby scarce resources---in this case, editor effort---must be efficiently allocated across various potential uses---in this case, the platform’s various articles. Editor effort has long been self-directed or coordinated via informal groups such as WikiProjects. It is not uncommon to see editors working alphabetically through categories of articles or choosing a topic such as naval history and systematically improving the content~\cite{wattenberg_visualizing_2007}.

Given the growing size and complexity of Wikipedia, however, there is a growing suite of automated recommender systems that help direct editor effort. Inherent in the design of these recommender systems are value-laden decisions about \emph{how} editor effort should be directed. SuggestBot~\cite{cosley_suggestbot_2007}, for example, provides personalized task recommendations based primarily on editors’ interests. This inherently foregrounds what individual editors themselves want to work on, which may differ from, for example, the types of content readers want to read. Warncke-Wang \emph{et al.}~\cite{warncke-wang_misalignment_2015} show that Wikipedia’s quality content already does not adequately meet reader demand for certain topics---e.g., LGBT issues---while providing a surplus of quality content for other topics---e.g., military history.

Directing Wikipedians towards articles that fit best with their individual interests may only exacerbate this misalignment. So, then, should editor effort be guided by reader demand instead? As a peer production platform, Wikipedia relies entirely on the efforts of volunteer contributors, so prioritizing reader interests at the expense of editor interests may do more harm than good, even for readers. As Warncke-Wang \emph{et al.}~\cite{warncke-wang_misalignment_2015} remark, \emph{``simplistic attempts to `force' volunteer contributors to work on high-demand topics rather than topics they find interesting and valuable may just cause them to leave or reduce their participation.''} Therefore, any prioritization scheme we use should, at the very least, be one that Wikipedia editors will actually \emph{want} to follow.

Zhu \emph{et al.}~\cite{zhu_value-sensitive_2018} emphasize that understanding and incorporating stakeholder values when designing algorithms makes them more likely to be well received by the community. We have already covered some ways in which two stakeholder groups---editors and readers---may have conflicting interests.
Further complicating the matter, the encyclopedia itself can be seen as a third stakeholder, with its own values and interests separate from those of the individual editors who produce its content.

Recent work in recommender systems (e.g., {~\cite{surer_multistakeholder_2018}}) has framed this problem as multi-stakeholder recommendation. While traditional recommender systems have tended to focus on the consumer, the multi-stakeholder framework expands the scope of recommender system algorithm design to incorporate and balance the interests of at least three primary stakeholder groups---consumers, providers, and the system~\cite{abdollahpouri_multistakeholder_2020}. In the case of Wikipedia, these might correspond to readers, editors, and the encyclopedia, respectively. This is a useful framework for the design of task-routing recommender systems on Wikipedia, because it acknowledges the potential for competing values between various stakeholder groups with regards to how content on the platform is prioritized.

In this paper, we take the encyclopedia itself as a stakeholder and interrogate its values with regards to how editor effort should be directed---or, in the language of multi-stakeholder recommendation, we identify \emph{the values of the system}. The non-hierarchical nature of Wikipedia makes this a difficult task, as there is no official governing body that functions as an authority with regards to the system's values. Therefore, in seeking to identify the encyclopedia's values, we build on prior work by Bryant \emph{et al.}~\cite{bryant_becoming_2005} that documents the ways in which experienced editors---i.e., ``Wikipedians''---begin to better understand and prioritize the integrity of the encyclopedia as a project over their personal interests. In other words, we identify the values of the system by interrogating those most familiar with and interested in forwarding the system's goals.

With this in mind, we examine \emph{Vital articles} (VA), a project on English Wikipedia where editors collaboratively come up with lists \emph{``to give editors guidance on which articles to prioritize for improvement.''} While VA’s process is editor-driven, its participants are attempting to focus not on their own interests but rather, according to VA’s FAQ page~\cite{noauthor_wikipedia_2020}, to \emph{``enhance the quality of the encyclopedia in the most essential areas.''} Through examination of VA participant demographics, we find that the vast majority of editors who engage in VA talk page discussions are highly experienced, and thus likely to be concerned with the needs of the encyclopedia as a whole. We begin by qualitatively analyzing editors’ public discussions on VA’s talk pages to reveal what Wikipedians perceive to be the values of English Wikipedia as an encyclopedia. We also conduct a smaller-scale analysis of a similar project on Meta-Wiki---\emph{List of articles every Wikipedia should have} \hl{(LoA)}---to further substantiate our results. We frame the following research question to guide our inquiry:

\begin{description}
\item [RQ1:] \textit{What criteria for directing editor effort emerge from English Wikipedia’s values?}
\end{description}

Our findings show that an English Wikipedia article’s priority is based on a combination of 1) its inherent importance and 2) its effects on the global composition of quality content on the platform. The latter type of consideration---which we call global criteria---highlights the need for evaluating recommender systems on Wikipedia based on the \emph{global} list of recommendations, rather than evaluating each set of recommendations in isolation, as is typical. Content balance---especially along gender and geographical lines---was one of the global criteria we observed. Wikipedia’s balance along these lines is of broad interest both within and beyond the community, and has been discussed extensively in the research literature (e.g.,~\cite{hargittai_mind_2015, antin_gender_2011, beytia_positioning_2020, bjork-james_new_2021}) and in the popular media (e.g,~\cite{baltz_wikipedias_nodate, cohen_define_2011, gleick_wikipedias_nodate, mandiberg_mapping_2020, torres_why_2016, resnick_2018_2018}).

We therefore take a quantitative approach to explore the issues of gender and geographical balance. Specifically, we examine \hl{VA alongside article importance metrics from prior work---PageRank, pageviews, and edit count---}and evaluate how using each of these four methods of prioritizing content would affect the gender and geographical balance of English Wikipedia. To aid in interpreting our results, we also examine whether there are any patterns in the topics that are surfaced most readily by each of these methods. We articulated additional research questions to guide this portion of our inquiry:

\begin{description}
\item [RQ2:] \textit{How do various prioritization methods compare in the topics they surface?}
\item [RQ3:] \textit{How do various prioritization methods compare in their representation of English Wikipedia’s values on balance?}
\subitem \textbf{RQ3a}: \textit{How do various prioritization methods compare in terms of gender representation?}
\subitem \textbf{RQ3b}: \textit{How do various prioritization methods compare in terms of geographical representation?}
\item [RQ4:] \textit{How would the composition of English Wikipedia’s highest-quality articles change with the use of different prioritization methods?}
\subitem \textbf{RQ4a}: \textit{How would the gender composition of English Wikipedia’s highest-quality articles change with the use of different prioritization methods?}
\subitem \textbf{RQ4b}: \textit{How would the geographical composition of English Wikipedia’s highest-quality articles change with the use of different prioritization methods?}
\end{description}

We now provide an overview of the relevant research literature in which we situate our work, followed by essential background information on VA.

\section{Related Work}

Prior work has examined ways in which recommender systems can be useful for directing editor effort on Wikipedia. For example, SuggestBot~\cite{cosley_suggestbot_2007}---a task-routing system that provides personalized article recommendations based on predicted user interests---was shown to increase editor content output by four times compared to providing random recommendations. Personalized recommendations in Wulczyn~\emph{et al.}~\cite{wulczyn_growing_2016} were shown to double editor engagement and substantially impact what content was edited. However, no research has focused on directing Wikipedians’ effort towards content that is higher priority \textit{for the encyclopedia}.

Previous work has shown how the deployment of algorithmic tools that are not sensitive to stakeholder values can cause unintended harms. Halfaker \emph{et al.}~\cite{halfaker_dont_2011}, for example, describe how automating the reversion of bad Wikipedia edits had the unfortunate consequence of driving away newcomers. This side-effect resulted from the design of algorithms that were blind to stakeholder values---in this case, Wikipedia’s ``please do not bite the newcomers'' policy~\cite{noauthor_wikipediaplease_2021}---despite meeting objective standards of performance such as accuracy.

The multi-stakeholder recommendation framework identifies three important stakeholders---the consumer, the provider, and the system~\cite{abdollahpouri_multistakeholder_2020}. In this paper, we focus on the system as stakeholder---i.e., we seek to understand \emph{English Wikipedia's} values with regards to recommender systems for directing editor effort. In doing so, we extend Zhu \emph{et al.}'s Value-Sensitive Algorithm Design (VSAD)~\cite{zhu_value-sensitive_2018}, which forwards the practice of incorporating stakeholder feedback early on in the design of algorithmic tools \emph{``with the goal of increasing stakeholder acceptance and engagement, reducing potential biases in design choices, and preventing compromise of important stakeholders' values.''} We view our qualitative inquiry as an implementation of VSAD's first step---\emph{understand stakeholders}---with a particular focus on the encyclopedia itself as stakeholder. By adopting this framework, we hope to increase the likelihood that recommender systems to direct editor effort are designed in such a way that would make them acceptable to English Wikipedia's community while successfully forwarding the encyclopedia's goals.

While we seek to understand the encyclopedia itself as stakeholder, it is obviously not possible for us to interrogate the encyclopedia in a literal sense. Instead, we structure our inquiry by building on Bryant \emph{et al.}'s{~\cite{bryant_becoming_2005}} findings about expert Wikipedia editors (emphasis ours):

\begin{quote}
    \emph{``For experts, or ``Wikipedians,'' the Wikipedia as a whole becomes more important than any single article or set of articles. Whereas initial edits tend to be focused on correcting individual articles, once users become Wikipedians, their goals expand. Although they continue to improve the quality of the content in individual articles, \textbf{their motivation seems to become rooted in a concern for the quality of the Wikipedia itself.} They also become concerned with improving the community. In the move from novice to Wikipedian, goals broaden to include growing the community itself and improving the overall quality and character of the site.''} 
\end{quote}

Because of their concern with the quality of the encyclopedia as a whole, expert Wikipedia editors---``Wikipedians''---are 
an excellent (though admittedly imperfect) source for understanding the values of Wikipedia as an encyclopedia, independent of the individual editors' interests. This leads us to use \emph{Vital articles} (VA)---a collaborative project mostly engaged in by expert Wikipedians---as a proxy for the encyclopedia's values with regards to article importance.

The concept of article importance has been used in prior work on underproduction in Wikipedia~\cite{gorbatai_exploring_2011}, and is relevant to consider in selecting articles towards which to direct editor effort. An informal review of Wikipedia article importance studies on Meta-Wiki~\cite{noauthor_researchstudies_nodate} finds that, while attempts to operationalize the concept of article importance are numerous and varied, PageRank is by far the most commonly used importance algorithm, appearing in 14 of the 50+ papers reviewed. Shuai \emph{et al.}~\cite{shuai_comparative_2013}, for example, use PageRank to compute and compare the scholarly and Wikipedia importance rankings of papers, authors, and topics in Computer Science. Lages \emph{et al.}~\cite{lages_wikipedia_2016} compare Wikipedia university rankings---also derived using PageRank---to a reputable human-curated university ranking. Similar methods are used in \cite{eom_highlighting_2013} to measure the importance of people across different Wikipedia language editions, \cite{eom_time_2013} to investigate the importance of Wikipedia articles over time, and \cite{eom_interactions_2015} to examine how Wikipedia and historians differ in their rankings of important historical figures.

Warncke-Wang \emph{et al.}~\cite{warncke-wang_misalignment_2015}, by contrast, take an approach to importance that centers reader demand. By examining the alignment of content quality (i.e., supply) and readership (i.e., demand), they find significant misalignment on the platform that systematically affects specific topics---e.g., articles on LGBT issues are too low quality (compared to reader interest), while articles on military history are too high quality (compared to reader interest). But despite concluding that Wikipedia does not adequately meet reader demand in these ways, they also question whether perfect alignment between content quality and readership is desirable at all. In other words, should we consider articles more important purely by virtue of their being more popular? If English Wikipedia as a platform has the goal of maximizing value for its readers, using reader demand---i.e., pageviews---as a measure of article importance might make sense.

In the quantitative portion of this paper, we examine how using these two aforementioned proxies for importance---PageRank and pageviews---to direct editor effort would compare with using VA. \hl{Though it is less common in prior research on importance, we also examine each article's total edit count, because it could reasonably serve as the basis for a standard recommendation engine---i.e., one that looks at what others have edited in the past to determine what users are likely to want to edit in the future.} More specifically, we analyze each prioritization scheme through the lens of the encyclopedia’s values on gender and geographical balance and investigate how adopting them might affect the composition of quality content on English Wikipedia. In the process, we keep in mind how demographics can potentially stand in the way of actualizing values. Lieberman and Lin~{\cite{lieberman_you_2009}}, for example, show that Wikipedia editors can somewhat reliably be tied to narrow geographic areas---usually their places of birth or residence---based solely on their editing histories. Hecht and Gergle~{\cite{hecht_measuring_2009}} similarly document the phenomenon of self-focus bias, where editors mostly contribute content that relates to themselves. We highlight places in our results where these phenomena seem to present themselves and explore their implications in the discussion section.

\section{Background: Vital Articles}

\emph{Vital articles} (VA) is a collaborative Wikiproject made up entirely of volunteer Wikipedia editors who are motivated to participate to forward its mission. Like other Wikiprojects on the platform, it does not have any official authority, but is rather a means by which editors organize their efforts towards a particular goal or set of goals---in this case, compiling lists of articles to be prioritized. From VA's home page:

\begin{quote}
    \emph{``Vital articles are lists of subjects for which the English Wikipedia should have corresponding featured-class articles. They serve as centralized watchlists to track the quality status of Wikipedia's most important articles and to give editors guidance on which articles to prioritize for improvement.''}
\end{quote}

Echoes of this mission statement exist in editors' discussions about VA's contributions to the encyclopedia. On April 25, 2020, an editor proposed the abolition of VA, citing its low activity and its conceptual overlap with LoA. Most of those who responded (7 of the 9) disagreed with the proposal, arguing that VA is \emph{``a useful device for establishing a hierarchy of subjects''} and that it helps \emph{``identify Wikipedia-wide articles that need attention,''} among other things. These justifications implicitly substantiate our earlier assertion that Wikipedians are subject to the economizing problem; a hierarchical list of articles that need attention is useful precisely because there isn't enough attention to go around. Many editors, moreover, see value in having a ranking that is specific to \emph{English} Wikipedia. (LoA is meant to be language-neutral.) One editor highlighted this using a concrete example:

\begin{quote}
    \emph{``we can see that the vital articles project has made the determination that Indo-European languages are a focal point at level 3 of vital importance, with only 3 non-Indo-European languages appearing - whereas this clearly would not be the case on, for instance, Chinese, Japanese or Arabic Wikipedia; for good reason, as on those wikis Indo-European languages are not likely to be as important a research interest as they are on Indo-European language wikis.''}
\end{quote}

We do not have a clear empirical measure of VA's actual impact on English Wikipedia---indeed, such an inquiry is well beyond the scope of this paper. However, we note that Wikipedians have engaged with questions about the project's merits and mostly agree about its overall usefulness as a tool for guiding effort, thereby facilitating more effective economizing. We also found examples of Wikipedians organizing contests and providing grants to improve the quality of articles in the VA list, on both English and Vietnamese Wikipedia.\footnote{\url{https://en.wikipedia.org/wiki/Wikipedia:The_Core_Contest}}\footnote{\url{https://meta.wikimedia.org/wiki/Grants:Project/Rapid/Improve_the_vital_articles_in_Vietnamese_Wikipedia}} Moreover, as one editor mentioned (and as we quantify below), VA participants themselves are highly active editors who \emph{``have recently (within the last few months, many of them within the last week) contributed to Wikipedia.''} Since VA participants are highly active editors, it is likely that their judgments about VA's usefulness also apply  to other such highly-active editors.

VA consists of five lists, each corresponding to a different level of ``importance,'' where importance is the desirability of bringing an article up to FA status. Level 1 contains the 10 most important articles, level 2 contains the top 100, level 3 contains the top 1,000, level 4 contains the top 10,000, and level 5 contains the top 50,000. There is no differentiation in rank between articles in the same level. As of July 2020, Wikipedians have so far filled in VA with 44,118 unique articles.

Decisions about what to include in VA happen through a discussion process on each level’s respective talk page, where users argue about and vote on proposed changes to the list. VA’s home page states that articles \emph{``should not be added or removed from this list without a consensus on the talk page,''} so the final decisions that are made about an article’s inclusion or exclusion from VA can reasonably be expected to reflect the general consensus of Wikipedians who are involved in the project. Perhaps more importantly for our work, analyzing VA’s rich talk page discussions can provide insight into the encyclopedia's values, as Wikipedians repeatedly reference them in arguing that an article should (or should not) be prioritized.

\subsection{Participant Demographics}

While VA is a convenient source of focused discussions between Wikipedians on how articles should be prioritized according to the encyclopedia's values, its participants are not a representative cross-section of all English Wikipedia editors, so it is worth going over their general demographics. Of the 794 editors we identify as ever having participated in VA talk page discussions, 79 have administrator privileges. For reference, only 1,119 accounts \emph{total} have administrator privileges on English Wikipedia. Even the non-administrators on VA are highly active and involved, however; VA participants have a median edit count of 6,419 on English Wikipedia---significantly higher than the median edit count of 2 for all registered accounts. While this highly-experienced editor population is more likely to espouse the encyclopedia's current values, it is certainly not representative. For this reason, we take their discussions as a source of values but do not focus on the amount of evidence for each value as an indication of its importance.

This more experienced editor population, perhaps unsurprisingly, skews heavily male. Of those who report a gender on their profile, only 7.1\% report they are women, compared to the 12.9\% who so identify on Wikipedia as a whole. This belies what is likely to be the actual gender distribution on VA, however, as men may be more likely to report gender on Wikipedia than women~\cite{hill_wikipedia_2013}. 40\% of VA participants report their gender as compared with only 2\% of Wikipedians overall, so the true difference in editor gender balance between VA and English Wikipedia as a whole is likely to be even more pronounced than our figures suggest. We later discuss how this lack of representation may cause the articles selected for inclusion by VA participants to be less gender-balanced than what English Wikipedia's values as espoused by this and other communities of editors would likely indicate.

LoA---which we use to substantiate our results from VA---is similarly composed of highly experienced editors. Of the 294 editors we identify as having participated in discussions on the project, 96 have most of their edits on the English language edition of Wikipedia. This subset of editors has a median of 1,289 edits on English Wikipedia, meaning they are also highly experienced ``Wikipedians''. Other represented language editions include Russian (28 editors), Swedish (16 editors), Catalan (11 editors), and Dutch (11 editors). Around 70 other languages are represented by fewer than 10 editors who primarily edit in those editions. Overall, those who have participated in discussions on this project have a median edit count of 10,713 across all Wikipedia language editions. This editor group's similarities to---coupled with its differences from---VA, make it a useful tool for substantiating the article prioritization values we extract. While this alternative group is also composed of highly experienced editors who are likely motivated to improve the quality of the project as a whole---rather than being solely concerned with their own interests---its differences from VA mean values that are expressed by both communities are likely to be values of the encyclopedia as a whole, rather than those of a specific project's members.

\section{Methods --- RQ1}

\begin{description}
\item [RQ1:] \textit{What criteria for directing editor effort emerge from English Wikipedia’s values?}
\end{description}

\subsection{Main Analysis}

Most of the VA discussions we analyzed were in a standardized VA proposal format, which includes a desired action in the title along with anywhere from a sentence to several paragraphs of justification for taking the action (Fig.~\ref{fig:vadiscussion}). Posting a proposal to one of the VA talk pages implicitly invites other Wikipedians to argue and vote on the proposed action, creating rich---and sometimes heated---back-and-forth conversations between highly opinionated editors.

We used thematic analysis to qualitatively study these VA talk page discussions. We first divided the discussions so that each distinct conversation about an article or set of articles would be analyzed as a whole. We stratified and sorted the data so that we would cycle through all 5 levels of VA equally as we went down the list of discussion content. For each sentence in each user comment, we first asked whether the editor contributes to discussion about VA beyond just indicating support for or opposition to a previously stated argument or proposal. This was intended to remove statements from consideration that did not provide insight into users’ reasoning---e.g., lines that merely read \emph{``Support.''} If the sentence contained potentially useful content, two researchers summarized each distinct statement made by the editor in the sentence and created an open code for it.

We began with 300 open codes---each corresponding to a distinct paraphrased editor statement, often a justification for or against a proposal. We used iterative thematic clustering to group the codes based on the justification criteria stated or implied in them. We reached data saturation---the same concepts kept coming up, with no new concepts emerging---well before clustering was over, so we stopped with the 300 open codes we had already generated.

\subsection{Robustness Check}

\hl{Afterwards, we performed the same open coding process on discussion data from a similar project---\emph{List of articles every Wikipedia should have}\footnote{\url{https://meta.wikimedia.org/wiki/List_of_articles_every_Wikipedia_should_have}} (LoA). LoA differs from VA in that it is not English Wikipedia-centric. It is a project on Meta-Wiki---a central platform for editors from the various Wikimedia projects and language editions to discuss Wikimedia project-related policy, analysis, coordination, and so on---that aims to provide a small list of 1000 important articles for new Wikipedia language editions \emph{``so that they will contain a minimum amount of basic, useful information.''} We used our analysis of LoA to check the robustness of our findings from VA and determine whether those results apply more generally than just VA.} We started with 100 open codes and clustered them on the same board as our open codes from VA. No new clusters had emerged by the time these additional 100 codes were clustered---we were still at data saturation, with all of the same concepts continuing to surface.

All results for RQ1 were derived from our \hl{analysis of VA and the subsequent robustness check on LoA.} We primarily report results from our analysis of VA as it is the more extensive and active project, but we \hl{also provide details of our findings from the LoA-based robustness check at the end of the Results section to show that our results describe the values of the encyclopedia---not just of VA's participants.}

\subsection{Privacy Considerations}

All analyzed discussion data from both projects are fully public. Despite this, we saw no compelling reason to include editors' usernames in this paper and therefore chose not to. Data were obtained from Wikimedia's database backup dumps{~\cite{noauthor_wikimediadump_nodate}}, thereby allowing ample time for editors to remove problematic or sensitive content they may have posted unintentionally.

\section{Results --- RQ1}

\begin{description}
\item [RQ1:] \textit{What criteria for directing editor effort emerge from English Wikipedia’s values?}
\end{description}

Our thematic analysis reveals several criteria Wikipedians commonly use in these conversations to argue for an article’s inclusion in---or exclusion from---the VA list (Table \ref{tab:criteria}). We separate these prioritization criteria into two broad categories:

\begin{enumerate}
    \item \textbf{Importance criteria}: those relating to an article’s importance based on its inherent characteristics.
    \item \textbf{Global criteria}: those relating to an article’s ability to promote or impede the encyclopedia's values with regards to the global composition of high-quality content.
\end{enumerate}

Some criteria similar to our own arise in the FAQ section for VA~\cite{noauthor_wikipedia_2020}, under a list of \emph{``commonly held notions''} that \emph{``have become prevalent''} in the project's discussions. We look for evidence to support these claims, but also unearth additional criteria---e.g., gender balance---through our more systematic qualitative inquiry. In the subsections that follow, we describe each prioritization criterion and category in detail, with supporting examples from VA talk page discussions.\footnote{Some editor quotes contain spelling and grammatical errors. None of these errors significantly alter legibility, so we do not correct or otherwise alter any cited quotes.}

\vadiscussion

\subsection{Importance Criteria}

Merriam-Webster defines \emph{important} as \emph{``marked by or indicative of significant worth or consequence.''} \cite{noauthor_definition_nodate} We adopt a similarly intuitive and broad definition of importance here. \textbf{Importance criteria are, simply, those that Wikipedians use to argue for an article’s inherent worth or value to the encyclopedia.} We begin by listing and describing all such criteria derived from our thematic analysis, with the help of examples from VA talk page discussions.

\subsubsection{Everyday Significance}

Topics that are relevant to everyday human life are widely considered important. For example, one VA contributor argued that the Math article should be included over Philosophy in level 1 because \emph{``I use pure math every day, but I don’t adhere to any particular philosophy.''} Similarly, another editor argued against the removal of Sleep from level 2 because \emph{``an activity that takes up 1/3 of your lifetime seems to be pretty vital to me.''}

\subsubsection{Cultural Significance}

Cultural significance may immediately bring to mind articles about cultural figures and celebrities, and it is certainly often applied to that category of article. Michael Jackson, for example, was proposed as a level-3 article in part because he \emph{``greatly influenced the development of music videos.''} Cultural significance is also, however, used as justification for the inclusion of Sun in level 2. \emph{``Sure, it’s a part of the Universe,''} one editor wrote, \emph{``but it has immense everyday and cultural significance.''} Another editor similarly argued for the addition of Sport to level 2 because \emph{``Sports have in some form been a part of the vast majority of cultures for much of there history.''}

\criteria

\subsubsection{Historical Significance}

A topic’s relevance to history is often used by Wikipedians as evidence of its importance to the encyclopedia. This applies perhaps most obviously to historical figures such as Aristotle (VA level 3), but also to concepts that have been consequential in the course of human history. One editor, for example, argued for the inclusion of Race in level 4 because \emph{``the concept has had a massive role in human history.''} In a different discussion, also in VA level 4, an editor argued in favor of keeping Bourgeoisie because it is \emph{``historically and culturally quite important''} despite being outdated and no longer relevant to modern political discourse.

\subsubsection{Enduring Significance}

Wikipedians repeatedly expressed a desire to avoid recency bias, so topics in VA are typically expected to have enduring significance throughout human history. Enduring significance underlies all other importance criteria that contain ``significance'' in their names. This does not, however, restrict the list to only covering topics that have been important for hundreds or thousands of years already---e.g., Aristotle, Music, or Clothing. Wikipedians might still determine that a fairly recent topic will have enduring significance due to the magnitude of its current significance. For example, in March 2020, a proposal to add 2019-20 coronavirus pandemic to VA level 4 passed unanimously with 15 votes. One editor expressed \emph{``strong support''} for the proposal, stating that \emph{``The world-wide impact of this event totally blows away any recency bias. The repercussions will be felt for many decades, at the very least.''}

It is worth briefly noting here that, while most historically significant topics might also be enduringly significant---seemingly by definition---the two criteria are distinct. For example, the argument we previously mentioned for Bourgeoisie relies on the concept's importance to history, while also acknowledging that it is not enduringly significant \emph{except} in its relevance to history---i.e., it is enduringly historically significant. By contrast, 2019-2020 coronavirus pandemic is not (yet) historically significant, but is expected to be enduringly significant to the everyday human experience---i.e., it has enduring everyday significance. In this latter example, we do not mean to imply that there is no case to be made for the coronavirus pandemic's historical significance, but rather to point out that there is an entirely distinct argument that can be made about its enduring significance without reference to its historical significance.

\subsubsection{Breadth}

Broader articles are considered more important than more specific articles. In some cases, an article’s breadth---or lack thereof---is used to argue that it belongs at a different level. For example, one editor opposed the addition of English Language to VA level 2 because it is \emph{``too specific for this level.''} In other cases, breadth is used as a general indicator of importance similar to the other criteria. For example, an editor supported swapping Emotion for Human Behavior at level 2 because the latter \emph{``seems to me broader''} and could therefore \emph{``describe broader issues such like politics.''} In a different discussion, an editor supported replacing Myth with Folklore at level 2 because \emph{``Folklore is the broader and more fundamental article.''} Another editor also supported this proposal \emph{``unless the myth article is renamed back to mythology and changes in scope.''}

\subsection{Global Criteria}

Our thematic analysis also reveals criteria---which we call global criteria---that stem from the values English Wikipedia has regarding VA's overall composition. Unlike importance criteria, applying global criteria requires considering characteristics of specific articles \emph{within} the particular context of VA as a whole. We once again list and describe all such criteria with examples from VA discussions.

\subsubsection{Balance}

Several Wikipedians expressed that VA should be balanced along a variety of dimensions. One editor, for example, opposed the addition of Universe to VA level 1 because, while the article is important, \emph{``it would only exacerbate the major problem the list has, which is the STEM bias.''}

Most of the balance concerns, however, centered around a desire to mitigate representational disparities in geography or gender. For example, one editor argued for the inclusion of Shaka (Leader of the Zulu Kingdom) in VA because \emph{``it is nice to have a fairly `global' coverage that isn't entirely Western-centric.''} Another editor noted \emph{``we have Characters from Western folklore; shouldn’t we also have Characters from Eastern folklore?''} In a discussion about whether to add Sport, a third editor argued that \emph{``if sport receives enough support then I think we should add an almost equivalent female dominated activity to balance things out (maybe dance).''}

\subsubsection{Non-redundancy}

Arguments based on redundancy are extremely common across VA discussions; a plurality of our open codes were assigned to this cluster. An article might fail the test of non-redundancy if the important aspects of its content are already covered by other articles in VA. For example, an editor opposed adding Universe to VA level 1 because \emph{``everything on Earth is covered by Earth, and everything beyond Earth is of interest pretty much only for astronomy, which is covered by Science.''} Another editor argued for the exclusion of Martin Luther King Jr.\ from VA level 3 because \emph{``as much as I respect MLK, Gandhi set the foundation for his ideals.''}

\subsubsection{Completeness}

Wikipedians commonly argue in VA for the inclusion of topic areas that should be covered in high quality by the encyclopedia. Unlike balance, completeness has to do with making sure a certain area of information is covered, rather than ensuring two or more areas are covered equitably. In the aforementioned discussion about whether to add Martin Luther King Jr.\ to VA level 3, one editor noted that \emph{``the only type of activism we lack is women's rights - of which i would support Emmeline Pankhurst.''} Elsewhere, another editor argued that \emph{``if we're not adding human behaviour we should definitely keep psychology''} in VA level 2 so that the important topic of human behavior remains covered. In a different discussion, an editor responded to a proposal for removing Book from level 2 with a potential compromise, saying \emph{``I think I’d like to see printing listed''} instead.

\subsection{Criteria in Conflict}

Multiple criteria are frequently used in the same discussion, often by different editors taking opposing positions on an article or set of articles. One such discussion, for example, begins with an editor proposing that Michael Jackson be added to VA level 3 because he \emph{``really represents the Pop music pinnacle. He also greatly influenced the development of music videos [...] and his legacy in Dance is widely known and timeless.''} In just these two sentences, the editor employs both cultural significance---citing Jackson’s influence on pop culture---and enduring significance---referring to the timelessness of his legacy---as criteria.

Another editor then employs an entirely different criterion---historical significance---to oppose the proposal. \emph{``We have fewer than 150 people at his level and I have a lot of trouble believing he’s one of the 150 people most vital to the course of human history,''} they reply. \emph{``Take the example of Donald Trump. Everyone knows him, to a greater extent than Jackson, but he isn't even in the top 2000 people.''} A third editor then brings in the global criterion of balance by adding that \emph{``I'd even put Édith Piaf over Jackson because we need a woman in music more than two male pop singers.''}

This example highlights that Wikipedians perceive the encyclopedia's various values as having different relative weights. In fact, some Wikipedians reference prioritization criteria---e.g., general reader interest---that did not even make it into our final list because they only surfaced once or twice by the time we had reached data saturation.\footnote{Notably, editor interest is not mentioned in VA discussions at all---i.e. editors do not justify an article's priority by their own interest in writing about it.} VA’s process, however, requires that a consensus be reached on any proposed changes, so we assume that VA mostly includes what its participants might collectively consider to be valuable to the encyclopedia, despite strong disagreements between editors on many of the particulars.

\subsection{Results of Robustness Check}

Analysis of LoA---a project on Meta-Wiki that aims to establish a list of highly important articles for new Wikipedia language editions to work on first---reveals the same criteria we outlined above. \hl{Discussions from LoA provide support for \emph{all} other criteria outlined above. The emergence of these same themes in a different project with prioritization-focused goals strongly suggests that the criteria we extracted are not specific to VA, but rather reflect the encyclopedia's general values on article prioritization.}

Interestingly, however, concerns about geographical balance were much more frequently expressed in this project. In one lengthy discussion, for example, an editor illustrated how bias can subtly creep in through the seemingly neutral topics editors decide the list should cover:

\begin{quote}
    \emph{``For example, for two of the most important cultural worlds, the Chinese (China, Japan, Corea) and the Muslim (Arab and Persian cultures) the most important art is not painting or even music, (arguably) it's calligraphy. There is not a single calligrapher in the list.''}
\end{quote}

The greater emphasis on geographical balance likely results from the project's goal as a resource for \emph{all} language editions of Wikipedia, in contrast to VA's focus on English Wikipedia specifically. Like VA, however, concerns about gender balance were also present here. In a different discussion, an editor highlighted how the demographics of Wikipedia editors can make it difficult to achieve completeness in a way that is not male-focused:

\begin{quote}
    \emph{``One one hand, I'm surprised it [Menstruation article] isn't here, but then as one of the x-deficient 90\% of editors, I wouldn't have even thought to add it.''}
\end{quote}

\hl{With these differences and---more importantly---similarities} in mind, we now move on to our quantitative study, in which we explore the outcomes of VA in comparison with other prioritization methods. Because we used a different set of methods to answer each of the three remaining RQs, we cover each RQ's methods and results together. First, however, we begin with descriptions of each of the four prioritization methods used.

\section{Prioritization Methods}

In the following sections of our paper, we examine four possible ways of generating a list of high-priority articles to be used by a recommender system to direct editor effort. \hl{Two of these methods---PageRank and pageviews---have been widely used in prior work as ways of approximating article importance on Wikipedia, and the other---edit count---could reasonably serve as the basis for a recommendation engine that looks at past edits to determine what users might want to edit in the future. We examine these metrics alongside VA, which serves as a proxy for the encyclopedia's values \emph{as implemented by Wikipedians}.}

\subsection{PageRank}

We use PySpark’s~\cite{noauthor_apachespark_2021} implementation of PageRank to apply the algorithm to the internal network of links between English Wikipedia articles on July 31, 2020 and using the settings from Thalhammer and Rettinger.~\cite{thalhammer2016pagerank} We exclude template links from the network to avoid strongly biasing PageRank towards articles that are automatically linked to from many articles---e.g., International Standard Book Number (ISBN).

\subsection{Pageviews}

We rank all articles from highest to lowest total user pageviews\footnote{\url{https://dumps.wikimedia.org/other/pageviews/readme.html}} from August 2019 to July 2020, inclusive (one year's worth of data).

\subsection{Edit count}

For edit counts, we consider only unique edits by non-bot registered users for each article as of July 2020.\footnote{\url{https://wikitech.wikimedia.org/wiki/Analytics/Data_Lake/Edits/MediaWiki_history}} We exclude bots and non-registered users because we want to understand where \emph{Wikipedians'} efforts have historically been directed.

\subsection{Vital articles}

We extract the 44,118 unique articles that were in VA as of July 31, 2020. In VA, higher level articles are also included in the lower levels---e.g., Mathematics is included in all VA levels---not just level 1---and level 5 contains all articles in VA. To better differentiate between the levels in our analyses, however, we assign each article only to its highest level---e.g., we consider Mathematics to only belong to level 1.

\section{Methods --- RQ2}

\begin{description}
\item [RQ2:] \textit{How do various prioritization methods compare in the topics they surface?}
\end{description}

WikiProjects are collaborative projects on Wikipedia that editors use to coordinate efforts on a particular set of articles, usually related to a single topic area. As such, they provide us with a convenient categorization scheme. We assume an article’s presence in a WikiProject indicates that it falls under the topic delineated by that WikiProject. For example, an article listed by WikiProject:Science is considered to be an article about science. \hl{Articles that relate to more than one topic can exist in multiple WikiProjects simultaneously.} We use normalized discounted cumulative gain (nDCG)---a common measure of search engine ranking quality---to determine how strongly each prioritization method represents different topics. For each WikiProject with at least 100 articles, we first calculate discounted cumulative gain (DCG) as follows:

\[ DCG = \sum_{i=1}^{p}rel_i/log_2(i+1) \]\

$p$ is the number of articles in the WikiProject and $i$ is the rank. $rel_i$ = 1 if the article at rank $i$ is in the WikiProject and 0 if it is not. IDCG is the ideal DCG---i.e., DCG if all of the top $p$ articles in the ranking are in the WikiProject. We then find nDCG, which is always between 0 and 1, by dividing the two:

\[ nDCG = DCG/IDCG \]\

For VA, we consider articles in the same level to be tied---e.g., all articles at level 1 are tied for rank 1, all articles at level 2 are tied for rank 11, and so on. \hl{nDCG for a WikiProject with 100 articles would therefore include all level 1 and level 2 articles in VA, and $rel_i$ at, for example, $i$ = 11 would correspond to the \emph{proportion} of articles at level 2 that are in the WikiProject.} 

We apply nDCG to each of our four prioritization methods---VA, PageRank, pageviews, and edit count---and report the top 10 WikiProjects for each. A WikiProject whose articles are generally at the top of a prioritization method's ranking will have higher nDCG than a WikiProject whose articles are lower in the ranking or missing from it altogether. In other words, a high nDCG indicates that the WikiProject is well-represented in the ranking so, for example, a high nDCG for WikiProject:Science in VA indicates that VA contains many highly-ranked articles related to science. We removed from consideration WikiProjects that do not correspond to particular topics---e.g., the Cleanup WikiProject, where editors focus on fixing formatting, spelling, and grammatical issues across various articles.

\ndcg

\section{Results --- RQ2}

\begin{description}
\item [RQ2:] \textit{How do various prioritization methods compare in the topics they surface?}
\end{description}

Based on nDCG results (Table \ref{tab:ndcg}), the top WikiProjects represented by VA pertain to core topics such as Anthropology, Philosophy, and Science, indicating that the Wikipedians who compiled VA are successful in ensuring broad and diverse topics are well represented in the ranking. Edit count’s top WikiProjects are wide-ranging, indicating a variety of areas that receive high editor attention. Edit count’s top WikiProjects are, however, narrower in scope---e.g., Rock music is near the top, as opposed to just Music more generally. So while the top most edited articles are varied in terms of topic area, editor focus is not centered on topics that are broad in the way that might be preferable for the encyclopedia.

Pageviews represents people and popular culture topics most strongly, which is consistent with the fact that \emph{popular culture} is in fact \emph{popular}, and pageviews is a reader-driven metric. We find WikiProject:Women in the list of best-represented WikiProjects. While this may be partially attributable to the fact that WikiProject:Women mostly contains biographies of women---and pageviews contains many biographies in its top 50,000 articles---Women also appears higher than WikiProject:Biography. We return to this and examine the actual gender-breakdown of each ranking method's biographies in the next section, showing that pageviews does in fact represent women more highly than any of our other prioritization methods.

PageRank, on the other hand, leans heavily towards countries and cities. This may result from the millions of town articles that have bot-generated links pointing to their corresponding countries, and from the many biographical articles that have links back to the person's city of birth~\cite{johnson_not_2016}. These consistent linking conventions likely cause countries to have out-sized PageRank scores compared to other types of articles. \hl{We explore this possibility in more detail in the Discussion section.}

Now that we have a general understanding of the topics surfaced most strongly by each method, we examine how they compare in terms of gender and geographical compositions.

\section{Methods --- RQ3}

\begin{description}
\item [RQ3:] \textit{How do various prioritization methods compare in their representation of English Wikipedia’s values on balance?}
\end{description}

VA groups articles into levels that grow in size exponentially. We accept this scheme at face value because it captures the intuition that the degree to which difference in rank is meaningful decreases as we go down any list of articles to be prioritized. For example, the difference between the 1st and 11th ranked articles is substantially more meaningful than the difference between the 1001st and 1011th. We therefore begin by separating the top 50,000 articles for each prioritization method into 5 levels based on the sizes dictated by VA (10; 100; 1,000; 10,000; and 50,000). Doing so also allows us to conveniently compare our other prioritization methods alongside VA.

To determine each prioritization method’s gender breakdown, we first filter its articles down to only biographies, as doing so gives us a clear and convenient way of assigning genders to those articles. We obtain gender for each article from Wikidata but only report statistics for cisgender men and women. There were fewer than 70 biographies of gender-fluid, non-binary, and transgender individuals for each prioritization method so it was not possible to determine statistically meaningful trends in their representation.\footnote{The proportions of biographies pertaining to gender-fluid, non-binary, and transgender individuals for each prioritization method were PageRank: 0.044\%, pageviews: 0.370\%, edit count: 0.210\%, and VA: 0.095\%. The proportion on English Wikipedia overall is 0.082\%.}

To determine geographical breakdown, we obtain country data from Wikidata, which allows us to assign each article to any countries it strongly pertains to.\footnote{We use any of the following properties for the Wikidata item associated with a given Wikipedia article: coordinates, country, place of birth, country of citizenship} The Wikimedia Foundation (WMF) uses the term Global North to refer to developed countries and Global South to refer to developing countries~\cite{noauthor_northsouth_2021}. WMF provides a list of each country’s classification among these two groups~\cite{noauthor_list_nodate}, which we use to classify articles as either pertaining to the Global North or Global South. We classify articles without geographical coordinate data---and those assigned to multiple countries spanning both regional classifications---as region-neutral.

Using these data, we examine and compare the gender and geographical breakdowns of articles surfaced by each prioritization method.

\section{Results --- RQ3}

\begin{description}
\item [RQ3:] \textit{How do various prioritization methods compare in their representation of English Wikipedia’s values on balance?}
\end{description}

\subsection{Gender Representation}

\begin{description}
\item [RQ3a:] \textit{How do various prioritization methods compare in terms of gender representation?}
\end{description}

We find that VA has no biographies at levels 1 and 2 (Fig. \ref{fig:rqthreegender}a) which, based on our qualitative results, is likely due to biographies being considered too specific for those levels. Pageviews and VA have a similar overall proportion of biographies---33.9\% and 33.4\%, respectively---though pageviews tends to rank biographies at higher levels than VA. PageRank has the fewest biographies across the board---only 13.5\%, with greater concentration at lower levels and, like VA, no biographies at levels 1 and 2.

Fig. \ref{fig:rqthreegender}b shows the gender breakdown of each ranking’s biographies, separated by level. 16.5\% of VA’s biographies are about women, which initially seems fairly close to the 18.5\% figure on English Wikipedia as a whole. Breaking it down further, however, reveals very low representation of women among biographies at higher levels. At level 3, we find that a mere 8.9\% of VA's biographies are about women, and level 4 is only slightly higher at 10.8\%.

Pageviews’ representation of women, on the other hand, is relatively high across the board, with 32\% of biographies in its top 50,000 representing women. Of the 3 biographies in level 1, all are about men, though they all pertain to people who had been in the news in the past year---Kobe Bryant, Donald Trump, and Jeffrey Epstein---indicating, unsurprisingly, that pageviews has a strong recency bias at the very top.

Edit count’s representation of women remains mostly stable and consistently above the overall proportion of women’s biographies on the platform. Though it does not contain as many biographies of women as pageviews overall, it sees a substantial spike to 32.1\% at level 2, putting it above pageviews at that level.

PageRank is consistently low in its representation of women, which is unsurprising given that the algorithm relies entirely on existing content to surface articles. Only 9.6\% of biographies in PageRank’s top 50,000 are about women, though that number is 11.6\% at level 3, meaning it still represents women more equally than VA at the very top of the rankings.

Overall, edit count and pageviews consistently represent women above the proportion of women’s biographies on the platform, while VA and PageRank both consistently represent them below that level. Pageviews leads the pack in representation of women rather consistently. PageRank is perhaps the worst at surfacing biographies of women in the aggregate, but still outperforms VA at the very top.

\rqthreegender

\rqthreegeo

\subsection{Geographical Representation}

\begin{description}
\item [RQ3b:] \textit{How do various prioritization methods compare in terms of geographical representation?}
\end{description}

56\% of VA’s articles are not associated with just one of the regions---e.g., the articles for Earth and Science are not specific only to countries in the Global North or Global South---and this number increases drastically by level; 51.7\% of articles at VA level 5 are region-neutral---very close to the platform-wide 50.7\%---while 100\% of articles at VA level 1 are, and the increase is consistent from one level to the next (Fig.~\ref{fig:rqthreegeo}a). This is consistent with our finding that VA participants see the inclusion of broader articles at higher VA levels as more consistent with the encyclopedia's values. Of our three metric-based rankings, edit count is the most region-neutral and PageRank is the least region-neutral, at all levels. PageRank \emph{is}, however, the only ranking that contains any Global South articles at level 1, though only a single one---India.

Fig.~\ref{fig:rqthreegeo}b shows that, among articles that are not region-neutral, VA also far outperforms all our metric-based rankings in representation of the Global South. VA level 2 contains only 5 non-region-neutral articles, however, all of which are continents---Africa, Asia, Europe, North America, and South America. The three metric-based rankings have similar geographical compositions at levels 4 and 5---between 15.7\% and 18.9\% Global South---but pageviews begins to skew heavily towards the Global North at level 3 (9.6\% Global South) and level 2 (9.8\% Global South).

Overall, VA represents region-neutral articles and articles relevant to the Global South at much higher rates than any of our three metric-based rankings. Among the metric-based rankings, edit count shows both the highest representation of region-neutral articles and the highest representation of Global South among non-region-neutral articles. PageRank is the least region-neutral across the board but, among non-region-neutral articles, represents the Global South better than pageviews at levels 1 through 4.

Using these same data, we now examine how each prioritization method, if adopted, might affect the composition of English Wikipedia's high-quality content.

\section{Methods --- RQ4}

\begin{description}
\item [RQ4:] \textit{How would the composition of Wikipedia’s highest-quality articles change with the use of different prioritization methods?}
\end{description}

Featured article (FA), Wikipedia’s highest article quality rating, is assigned based on strict criteria~\cite{noauthor_wikipediafeatured_2021}; as of July 2020, only 5822 of English Wikipedia’s 6 million+ articles are listed as meeting the bar for FA status~\cite{noauthor_wikipediafeatured_2020}. By examining the composition of FA, we can understand what types of information on English Wikipedia are available to readers in well-written, comprehensive, well-researched, neutral, and stable articles with good style~\cite{noauthor_wikipediafeatured_2021}. Moreover, English Wikipedia's Main Page, which gets approximately 15 million hits per day, features a short summary of one FA article every day~\cite{noauthor_wikipediatodays_2021}, so the composition of FA also affects what many Wikipedia readers get exposed to. It is therefore important to understand FA’s current composition and how various prioritization methods might affect it.

Between July 31, 2015 and July 31, 2020, the number of FA articles on English Wikipedia increased by 1257. We assume this trend will continue and use this to project forward 5 years to 2025. We ask: assuming new FA articles are added at the same rate, but in the order dictated by each of our prioritization methods, how would the composition of FA articles on English Wikipedia change? More specifically, what would happen to the composition of FA if we only bring articles from a lower level to FA after all articles at higher levels are at FA status? We use the same methods as for RQ3 to split articles for each prioritization method into levels and re-use the same gender and geographical classification schemes for articles.

\rqfourgender

\rqfourgeo

\section{Results --- RQ4}

\begin{description}
\item [RQ4:] \textit{How would the composition of Wikipedia’s highest-quality articles change with the use of different prioritization methods?}
\end{description}

\subsection{Changes to Gender Composition}

\begin{description}
\item [RQ4a:] \textit{How would the gender composition of Wikipedia’s highest-quality articles change with the use of different prioritization methods?}
\end{description}

Given identical gender distributions, methods that surface more biographies will have larger impacts on the gender distribution of FA’s biographies. Pageviews’ strong representation of women coupled with its high number of biographies therefore gives it a substantial effect on FA’s gender distribution, increasing the proportion of women from 15.4\% to 21.4\% over 5 years (Fig.~\ref{fig:rqfourgender}). The other methods' effects are much more modest. Edit count has a slight positive effect on women’s representation in FA biographies, increasing it to 16.4\%. PageRank decreases the representation of women by less than half a percentage point, to 15.1\%.

VA, meanwhile, has the strongest negative effect on women’s representation, bringing it down to 14.7\% of FA biographies. VA’s effect on the representation of women would begin to reverse once editors reached level 5 but, at the current rate, reaching level 5 would take approximately 4 decades. Given that Wikipedia has a constant stream of important new information to document at high quality---e.g., as previously mentioned, 2019-20 coronavirus pandemic was added to level 4 in 2020---we find it unlikely that the higher levels would stay static for long enough to ever necessitate that editors use level 5 to prioritize.

\subsection{Changes to Geographical Composition}

\begin{description}
\item [RQ4b:] \textit{How would the geographical composition of Wikipedia’s highest-quality articles change with the use of different prioritization methods?}
\end{description}

We find that all prioritization methods would increase the Global South’s representation in FA by a modest amount (Fig. \ref{fig:rqfourgeo}). We notice that the biggest changes, however, are in the proportions of articles that are region-neutral. Using PageRank to prioritize, for example, would increase the number of Global South articles the most, bringing their representation up to 5.7\% of FA articles from 3.9\%, but this increase comes with a substantial decrease in the proportion of articles that are region-neutral, from 58.9\% to 54.4\%.

Using VA, by contrast, brings the proportion of region-neutral articles up to 63.1\%. And while VA’s 4.4\% Global South representation seems lower than PageRank’s 5.7\%, VA also results in a lower Global North representation than any of the other ranking methods. Overall, the differences in the Global-North-to-Global-South ratios of the various prioritization methods are very small, and all four would increase the Global South’s representation in FA articles. VA, however, stands apart as the only prioritization method that also increases the proportion of articles that are region-neutral.

\section{Discussion}

Our results have several implications for the design of recommender systems to direct editor effort. We begin by discussing how our findings further reinforce the need for multi-stakeholder recommendation in task-routing for Wikipedia. We then show how the existence of global criteria reveals the need to consider \emph{aggregate} diversity as a key performance metric in this context. We provide recommendations for operationalizing importance in the context of directing editor effort, observe how editors' demographics can (and seemingly do) impede the encyclopedia's values with regards to article priority, and conclude with a brief discussion of this work's main limitations.

\subsection{Task-Routing on Wikipedia Needs to be Multi-Stakeholder-Focused}

Our results show several ways in which editors' own interests differ from those of the encyclopedia as a whole. More importantly, we see that there are inherent tensions and conflicts between their respective interests. \emph{Individually}, editors may just want to work on the things they find most interesting. There already are recommender systems for Wikipedia---e.g., SuggestBot~\cite{cosley_suggestbot_2007}---that help connect editors with articles based on their individual interests. \emph{Collectively}, however, Wikipedians may also want the community as a whole to prioritize certain articles based on their perceptions of the encyclopedia's values. Prior work has already shown that, in the aggregate, Wikipedians have greater interest in some topics over others, so personalization---i.e., fulfilling editors' \emph{individual} goals---may end up hindering balance, non-redundancy, and completeness---some of the \emph{collective} platform goals we identify.

The existence of these tensions further highlights the need for the multi-stakeholder recommendation framework in the design of task-routing recommender systems for Wikipedia. Interestingly, a recommender system that focuses only on the individual provider (i.e., editor) without regards to the system (i.e., encyclopedia) values is unlikely to fully satisfy even many individual editors---especially Wikipedians, who also have an interest in forwarding the encyclopedia's values. Our results make it clear that, perhaps unintuitively, building task-routing recommender systems that satisfy \emph{only} editors' values still requires taking a multi-stakeholder approach because of editors' concern with the encyclopedia's values.

In a literature review of multi-stakeholder recommendation, Abdollahpouri \emph{et al.}{~\cite{abdollahpouri_multistakeholder_2020}} describe several algorithmic approaches, some of which directly incorporate multiple objectives into the algorithm's loss function, and others which start with standard recommendations and then modularly re-rank them to take additional stakeholders into account. Our work does not inherently make clear which of these many approaches would be most appropriate for Wikipedia---and the literature has not yet settled on a set of ``gold-standard'' methods{~\cite{abdollahpouri_multistakeholder_2020}}---but it \emph{does} reveal the necessity of applying the multi-stakeholder recommendation framework to the problem of task-routing in this context.

\subsection{Global Criteria Reveal the Importance of Aggregate Diversity}

The relevance of global composition to VA---\emph{``lists of subjects for which the English Wikipedia should have corresponding featured-class articles''}---reflects the relevance of global composition of high-quality content to English Wikipedia as a whole. Our qualitative inquiry therefore reveals that an article's priority for English Wikipedia is partially a function of the ways in which it affects the global composition of quality content on the platform.

Prior work in recommender systems on Wikipedia has mostly evaluated the quality of recommendations based on each set of recommendations in isolation~\cite{herlocker_evaluating_2004}. Cosley \emph{et al.}~\cite{cosley_suggestbot_2007}, for example, evaluate Wikipedia's SuggestBot by asking users to each rate a single list of recommendations and then aggregating those ratings. In the context of recommender systems for directing editor effort on Wikipedia, however, similar evaluation strategies would be incomplete. To understand how well a recommender system forwards the encyclopedia’s values on global composition, we must shift our focus to the \emph{global list} of recommendations provided to all editors. For example, at the extreme end, a recommender system that provides a single user with exclusively military history-related articles that they consistently find interesting is not necessarily bad, but one that provides \emph{all} users with exclusively military history-related articles is unlikely to forward any of the global composition values we discovered.

These global considerations can be encapsulated by the concept of \emph{aggregate diversity} in recommender systems. In contrast to \emph{individual diversity}, which is satisfied by diversifying a single user's recommendations, aggregate diversity refers to the diversity of all recommendations presented to all users. Prior work by Mansoury \emph{et al.}~\cite{mansoury_fairmatch_2020} provides a useful blueprint for achieving higher aggregate diversity in practice. They begin by generating more recommendations than will eventually be presented to the user, then use a graph-based post-processing algorithm to identify a subset of \emph{``high-quality items with low visibility''}, which constitutes the final set of recommendations. By testing their algorithm on two datasets, they show that higher aggregate diversity can be achieved with comparable recommendation accuracy. Adapting similar algorithmic approaches to the problem of task-routing on Wikipedia would make it possible to directly incorporate the encyclopedia's values on global composition while still providing recommendations that are interesting to the particular editor.

We also know from prior work that editors are more likely to act on recommendations related to topics that meet their individual interests~\cite{cosley_suggestbot_2007,wulczyn_growing_2016}---i.e., a single recommendation for a military history article may be more effective on English Wikipedia than a single recommendation for an article related to LGBTQ+ issues if there is higher aggregate editor interest in the former topic. It is therefore not sufficient for the recommendations themselves to mirror English Wikipedia's values regarding content composition---they must also affect editor attention and effort in ways that create outcomes which actualize these values.

Recent work by Diaz \emph{et al.}~\cite{diaz_evaluating_2020} takes steps towards tackling this problem. They point out that standard evaluation methods focus on whether items in a ranked list are in the ``correct'' order without accounting for how much attention the items actually get. They present the concept of expected exposure---\emph{``the average attention ranked items receive from users over repeated samples of the same query''}---and propose that recommender systems should ensure items with equal relevance receive equal expected exposure. In the context of task-routing on Wikipedia, this would mean that merely balancing the number of recommendations between articles about, for example, military history and LGBTQ+ issues would not be sufficient, as military history articles might still receive disproportionate attention/exposure due to differences in editor interest. Task-routing recommender systems on Wikipedia should therefore incorporate Diaz \emph{et al.}'s insights to create systems that \emph{actually forward} stakeholder values instead of merely providing recommendations in an order that is \emph{consistent with them}.

\subsection{Evaluating Proxy Metrics for Operationalizing Importance}

In a recommender system for directing editor effort, we could consider an article’s relevance as analogous to some combination of its inherent importance and its interest to the particular editor---i.e., all global considerations being equal, we want to surface more important articles that the editor would find interesting to work on. Methods of operationalizing editor interest are already well established~\cite{cosley_suggestbot_2007}. Prior work has operationalized Wikipedia article importance in various ways, two of which---pageviews and PageRank---we explore in this paper. Our results reveal several ways in which these proxy metrics would not adequately forward the encyclopedia's values with regards to global composition; using PageRank to prioritize would result in less gender balance, while using pageviews would do the least to ameliorate geographical balance, neither of which is desirable. Even if we put global considerations aside and only consider articles' inherent importance, however, our results on topics surfaced show that both methods fall short in different ways when applied as relevance metrics in this particular domain. \hl{These shortcomings also highlight that the use of these metrics could not be justified by reference to stakeholder values---i.e., despite pageviews being reader-driven and PageRank being editor-driven, neither of them seems to capture those stakeholders' values in a meaningful way.}

\subsubsection{Pageviews as Proxy for Importance}

Pageviews does not explicitly communicate what readers believe should be prioritized, a fact that is substantiated by our results showing that pageviews most heavily surfaces articles related to popular culture. Prior work by Singer \emph{et al.}~\cite{singer_why_2017} highlights that \emph{``Wikipedia is read in a wide variety of uses cases''} ranging from simple boredom-motivated browsing to work-related research. The skew towards popular culture shows that pageviews likely favors interest- or boredom-driven reader use cases over work-related ones. It is not difficult to imagine that readers might care substantially less about content quality in these less ``serious'' use cases, even though they are more frequent as indicated by pageviews. \hl{While this statement is purely \emph{speculative} based on what we as researchers might reasonably guess readers value, it illustrates that} a reader who views two articles equally is not necessarily indicating that their quality is of equal importance to them. Readers are also likely influenced by the access methods---e.g., Google search accounts for the vast majority of visits to Wikipedia~\cite{mcmahon_substantial_2017}, articles featured on Wikipedia’s Main Page get a huge boost in pageviews~\cite{noauthor_wikipediapageview_2021}. It is therefore worth questioning whether pageviews can truly approximate reader values in a meaningful way, \hl{despite its usage in prior work as an importance metric that is guided by reader demand (e.g., ~\cite{warncke-wang_misalignment_2015})}.

\subsubsection{PageRank as Proxy for Importance}

As the most commonly used article importance algorithm, PageRank is the other obvious candidate. Our results reveal ways in which it is also unsuitable for determining article importance, however. PageRank’s original use was in surfacing highly authoritative web pages. When website A links to website B, that signifies an endorsement of website B by website A; PageRank, as originally conceptualized, uses this fact about linking conventions on the web to find content that is highly authoritative. By contrast, a link from article A to article B on Wikipedia does not signal any type of endorsement, as the linking conventions on Wikipedia are very different. For example, the fact that Hannah Glasse (18th century English cookery writer) links to the article on Gelatin dessert could not reasonably be understood as an endorsement of Gelatin dessert by the authors of Hannah Glasse’s article. Rather, it is merely a result of Gelatin dessert’s relevance to Hannah Glasse, who was the first to write about it in a cookbook.

This difference in linking conventions manifests clearly in our results on PageRank’s most surfaced topics; there is a convention on Wikipedia that every town or city links to its country, and every biography links to the person’s place of birth. This causes articles about locations to have lots of in-links, which in turn causes them to rise to the top of the PageRank ranks. PageRank, therefore, is more reflective of how Wikipedians conventionally link articles together than a proxy for authoritativeness or importance in any real sense. \hl{Furthermore, editor-driven importance metrics like PageRank (and edit count) may rely not only on editor values and interests, but also outside factors such as feasibility---e.g., the existence of suitable templates or the availability of acceptable sources~\cite{berson_reliable_nodate}.} This substantiates prior work by Hanada \emph{et al.}~\cite{hanada_how_2013} that finds only a moderate correlation between PageRank and Wikipedians' article importance assessments in WikiProjects.

\subsubsection{Broader Challenges in Operationalizing Importance}

Along with pageviews, they also have shortcomings that are likely to exist for many other metrics we might attempt to use as implicit proxies for article importance. In an effort-directing recommender system we are again concerned with importance in the sense of \emph{needing to be prioritized}. Recommendations should therefore be concentrated on low-quality items where work is needed (and improvement would matter, because it advances relevant values). As a result, the most relevant items are also those for which we likely have the least data available, further confounding the challenge of identifying important content, and essentially putting the recommender system in a perpetual cold-start scenario. Another way of putting this is that we are concerned with the \emph{ideal} states of items rather than their current states. Hence, rather than using implicit indicators of importance such as pageviews or PageRank, we need explicit ratings of importance assigned by Wikipedians who have knowledge of the desired ideal states, which takes a lot of time to crowd-source (as evidenced by the fact that VA is \emph{still} incomplete).

Crowd-sourcing in this way would severely limit recommender systems’ adaptability. Given the dynamism inherent in being an \emph{online, open-source} encyclopedia, we were somewhat surprised that VA participants put so much emphasis on avoiding recency bias and only chose to prioritize articles that are enduringly significant. This can be explained, however, by observing that while Wikipedia itself is highly dynamic, VA is not; it is compiled through a highly involved and slow manual process. The emergent constraints against recency bias may in fact be nothing more than a necessary response to the slowness of this process, rather than a reflection of the encyclopedia’s values. In other words, VA participants likely realized early on that an article that should be prioritized \emph{right now and only right now} may not be suitable for addition to a prioritization list that does not change quickly. Any methods of crowd-sourcing explicit article priority assessments from Wikipedians should be careful to avoid designs that impose similar time-based (or other arbitrary) constraints.

\subsection{Demographics Can Impede Community Values}

In principle, we encourage the use of explicit importance assessments to inform the future direction of editor effort. In practice, however, our results show that there are ways in which Wikipedians’ explicit article priority assessments can end up hindering the very values they intend to forward. 

\subsubsection{Gender Bias}

With regards to gender, we see a stark contrast between allocation of editor attention as measured by PageRank and as measured by edit count. Using PageRank as a prioritization method would lead to a decrease in gender parity over time, suggesting that articles on topics \emph{related to} women have historically been neglected by Wikipedia's editors. Examining edit count data, however, reveals that accentuating historical editing trends would still increase the proportion of high-quality \emph{biographical} articles about women over the next few years. This is an encouraging trend, given the wide-spread criticism of Wikipedia's gender gap, especially with regards to biographies~\cite{noauthor_criticism_2021, resnick_2018_2018, schlanger_wikipedia_nodate, cohen_define_2011, noauthor_wikipedias_nodate-1}.

Our results, however, show that using VA to direct editor effort would likely \emph{reverse} this trend. We were surprised to find such a sharp drop in the representation of women among VA’s higher level biographies, especially given our discovery of gender balance as one of the encyclopedia's values in our qualitative study of VA. These results, however, are consistent with Hecht and Gergle’s findings of self-focus bias on Wikipedia~\cite{hecht_measuring_2009}---i.e., that Wikipedians contribute more information that is important and relevant to themselves. Of the 794 editors we identified as having participated in VA talk page discussions at least once, only 7.1\% (56) of those listing a gender identify themselves as women, as compared to the 12.9\% who so identify on English Wikipedia as a whole. Furthermore, for an article to be included in VA, an editor must take the initiative to create a decently compelling proposal for it. This barrier to entry may exacerbate self-focus bias in the context of VA because editors are more likely to go through the trouble of introducing proposals for articles relating to topics they pay closer attention to, despite their ideological support for gender balance in the abstract. 

This lends additional credibility to Menking and Rosenberg's feminist critique of Wikipedia's epistemology~\cite{menking_wpnot_2021}. In evaluating Wikipedia's third pillar---\emph{``Wikipedia is free content that anyone can use, edit, and distribute''}---they state that \emph{``though Wikipedia may be explicitly endorsing the value or necessity of broad and diverse participation, its implementation of that value in its culture of participation may very well work at cross-purposes.''} Similarly in VA, though Wikipedians may explicitly endorse the encyclopedia's values on gender balance, the demographics of its participants may in practicality hinder the actualization of those values. We therefore support Menking and Rosenberg's revised version of the third pillar---\emph{``The integrity of Wikipedia is a function of the size and breadth of its community.''} To forward the encyclopedia's values with regards to the composition of quality content, we should aim to diversify the platform’s editor base so as to better equip them to achieve gender balance in a way that is consistent with those values.

\subsubsection{Geographical Bias}

Interestingly, we find that Wikipedians who participate in VA are able to transcend self-focus when it comes to geography. English Wikipedia's preference for broad articles over specific ones seemingly leads VA participants to represent region-neutral topics more than the alternative methods, and VA is successful in having a greater representation of the Global South than even reader demand or current editor attention allocation might lead us to believe is called for. This is despite editors being heavily concentrated in the Global North; a 2011 Wikimedia Foundation survey of Wikipedia editors~\cite{noauthor_editor_nodate} finds that \emph{``the only country from the Global South among the top 10 is India, with 3\% of survey participants listing India as their country of residence.''} This may point to gender being a bigger driver of self-focus for editors on Wikipedia than geography. 

Alternatively, this could be due to the focus on \emph{enduring} significance and differences in the degrees to which power has historically varied along geopolitical versus along gender lines. While places in the Global North do currently have more power---and therefore perceived importance---than those in the Global South, the geopolitical landscape has varied substantially throughout history~\cite{AssaJacob2015GDPG}. Egypt, for example, is classified as Global South irrespective of Ancient Egypt's substantial political and cultural power many centuries ago. By contrast, women have consistently been underrepresented among those in power throughout all of history and all or most civilizations~\cite{epstein_great_2007}. Thus, when looking at long-term/enduring significance, historical power dynamics may create stronger filters along the lines of gender than of geography. While we find this plausible as a partial explanation, however, more specific study of this is necessary to draw any firm conclusions.

\subsubsection{The Consequences of Demographics Apply to Readers Too}

Diversifying Wikipedia’s editor base is likely to serve readers’ interests better too, as our results show that they are \emph{also} self-focused. 30\% of English Wikipedia's daily readers---the primary drivers of pageviews---are women~\cite{johnson2020global}, as compared with an estimated 16\% of editors~\cite{hill_wikipedia_2013} and 7.1\% of editors who participated in VA talk page discussions at least once. When we conceptualize prioritization by reader demand as also giving a greater voice to the interests of women who interact with Wikipedia, our finding that using pageviews as a prioritization method would create drastically more gender parity in the platform’s quality content as compared with other methods make perfect sense. Our results show that readers are geographically self-focused as well; consistent with pageviews’ relatively high representation of the Global North is the fact that approximately 78\% of pageviews to English Wikipedia in July 2020 came from countries in the Global North, according to Wikimedia Statistics~\cite{noauthor_wikimedia_nodate}.

\subsection{Limitations}

We now discuss our work's main limitations, some of which we have already alluded to. In doing so, we hope to highlight a couple important gaps in knowledge that remain for future work to fill.

First, like most other qualitative work, the population of editors we use to ascertain English Wikipedia's article priority values is not representative of the community as a whole. English Wikipedia hosts a diverse set of experienced editors who participate across thousands of WikiProjects, the vast majority of which we do not examine here in detail. Thus, even though we are able to identify and substantiate the article prioritization criteria that are widely used by experienced editors to forward the encyclopedia's values, our methods preclude us from making statements about the relative weights of the criteria. To better understand how the broader English Wikipedia community weights the importance of each criterion, one would need to use, e.g., large-scale surveys of Wikipedia editors to obtain a more representative sample.

On the quantitative side, several of our analyses make simplifying assumptions. For example, for the purposes of evaluating each prioritization method's effects, we assume it is followed \emph{perfectly} by editors. While this provides insight into each method's particular biases, it does not paint a fully accurate picture of what the \emph{actual} effects of adopting each method would be. (This, in large part, depends on factors that are not intrinsic to the prioritization methods themselves---e.g., level of community adoption.) This highlights the need for future work that actually deploys and evaluates multi-stakeholder task-routing tools that incorporate our insights.

\section{Conclusion}

By studying article priority through the lens of English Wikipedia's values, we show that existing approaches for determining article importance fall short in many substantive ways. We uncover several of the challenges inherent in using recommender systems to automate the direction of editor effort on Wikipedia, and provide guidance for future work that centers the encyclopedia's values. Finally, we demonstrate that the community's gender demographics can prevent it from actualizing the encyclopedia's values, lending further credibility to feminist critiques of Wikipedia.

\bibliographystyle{ACM-Reference-Format}
\bibliography{ref}

\received{April 2021}
\received[revised]{November 2021}
\received[accepted]{March 2022}

\end{document}